\documentclass[sigconf]{acmart}
\AtBeginDocument{%
  \providecommand\BibTeX{{%
    \normalfont B\kern-0.5em{\scshape i\kern-0.25em b}\kern-0.8em\TeX}}}

\setcopyright{acmcopyright}
\copyrightyear{2018}
\acmYear{2018}
\acmDOI{10.1145/1122445.1122456}

\acmConference[Woodstock '18]{Woodstock '18: ACM Symposium on Neural
  Gaze Detection}{June 03--05, 2018}{Woodstock, NY}
\acmBooktitle{Woodstock '18: ACM Symposium on Neural Gaze Detection,
  June 03--05, 2018, Woodstock, NY}
\acmPrice{15.00}
\acmISBN{978-1-4503-XXXX-X/18/06}

\begin{document}

\title{How Can I Swing Like Pro?: Golf Swing Analysis Tool \\for Self Training}

\author{Chen-Chieh Liao}
\affiliation{%
  \institution{Tokyo Institute of Technology}
  \city{Tokyo}
  \country{Japan}
}
\email{liao.c.aa@m.titech.ac.jp}

\author{Dong-Hyun Hwang}
\affiliation{%
  \institution{Tokyo Institute of Technology}
  \city{Tokyo}
  \country{Japan}
}
\email{hwang.d.ab@m.titech.ac.jp}

\author{Hideki Koike}
\affiliation{%
 \institution{Tokyo Institute of Technology}
  \city{Tokyo}
  \country{Japan}
}
\email{koike@c.titech.ac.jp}

\renewcommand{\shortauthors}{Chen-Chieh Liao, Dong-Hyun Hwang and Hideki Koike}

\begin{abstract}
In this work, we present an analysis tool to help golf beginners compare their swing motion with experts' swing motion.
The proposed application synchronizes videos with different swing phase timings using the latent features extracted by a neural network-based encoder and detects key frames where discrepant motions occur. 
We visualize synchronized image frames and 3D poses that help users recognize the difference and the key factors that can be important for their swing skill improvement.
\end{abstract}



\begin{teaserfigure}
  \includegraphics[width=0.85\textwidth]{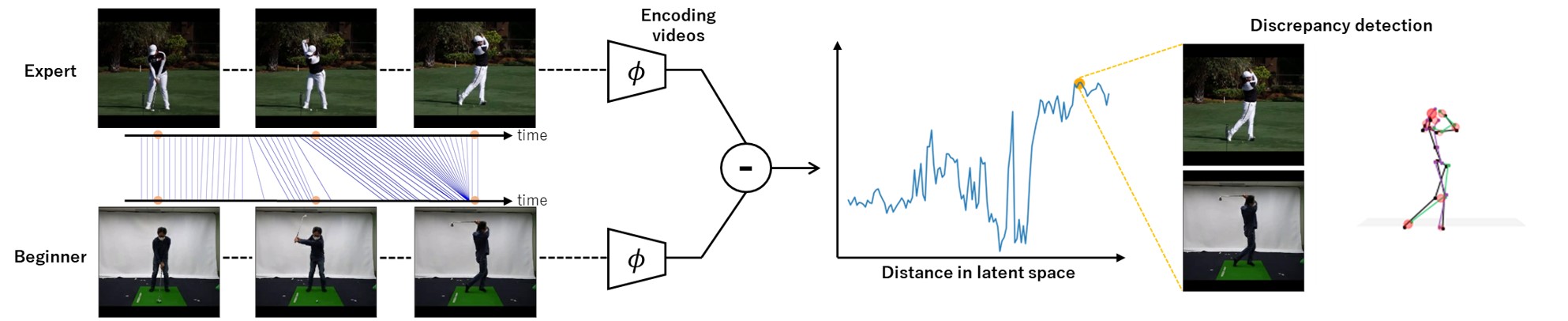}
  \centering
  \caption{Overview of the proposed approach and visualized analysis results.}
  \label{fig:teaser1}
\end{teaserfigure}

\maketitle

\section{Introduction}
In golf, correct swing forms lead to excellent outcomes. 
For golf beginners, a common way to understand and learn a correct swing motion naturally is to observe professional players' swings and imitate their forms. 
However, it is hard for beginners to specify key frames that they should focus on and which part of the body they should correct due to the inconsistent timing and the lack of knowledge.


In this work, we propose a golf swing analysis tool using deep neural networks to help the user recognize the difference between the user's swing video and the expert's one from the Internet or broadcast media.
The tool visualizes the image frames and human motion where the discrepancy between the expert and the user's swing is large, and the user can easily grasp a motion that needs to be corrected.

\begin{figure*}[h]
  \centering
  \includegraphics[width=0.85\textwidth]{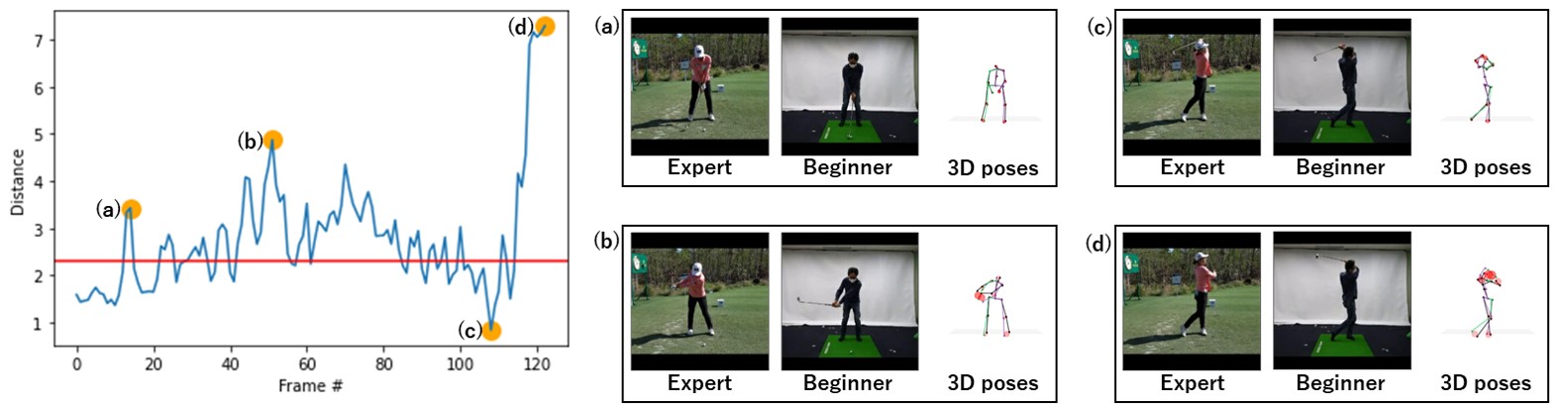}
  \caption{Comparison among the distance in the latent space, the two input images and the estimated 3D skeletons. The red line in the graph indicates the threshold for discrepancy detection. The colored skeleton and black skeleton indicate the user's pose and expert's pose, respectively. The intensity and radius of red circles indicate the degree of joint position difference of the two skeletons.}~\label{fig:teaser2}
\end{figure*}

\section{OUR APPROACH}
We create a prototype application that visualizes the distance of the latent space of two input videos, detects discrepant frames with adaptive threshold, and compares the detected frames with 3D human poses (Figure~\ref{fig:teaser1}).
Since the phasing and timing of a golf swing can vary from person to person, we synchronize the input video clips by implementing the Temporal Cycle-Consistency (TCC) network~\cite{TCC}. 
The self-supervised learning algorithm, which is used in the TCC, can be trained with unlabeled data such as videos on the Internet and broadcast, and so can generally fit many other types of motion analysis.

The inputs for the network are two video clips, as the outputs of the network are two compressed latent vectors. 
A distance between two latent vectors indicates the similarity of the two videos.
This characteristic of the latent space can be used for synchronizing videos within the same category of motion, and in this work, we focus on the distance between two videos in the latent space for detecting and retrieving the fine-grained discrepancy to compare between two different swing forms, especially between beginners and experts.

For the application implementation, we first extract the latent vectors frame by frame, which is the output of the TCC network, and then synchronize the input videos by calculating the Euclidean distance in the latent vectors. At this point, similar motions should appear to be close in the latent space; however, repeated motions appear frequently during a golf swing, i.e. a particular posture can be observed both during the backswing and downswing phase.
Therefore, we apply the Dynamic Time Warping algorithm to penalize the distance in the latent space to gain a better alignment.

To visualize the 3D poses of players, we use the HRNet~\cite{Hrnet} and a simple linear network~\cite{3dposebaseline}, then we applied Procrustes analysis to align two joint sets from different camera positions. This feature allows the users to see the motion difference between themselves and the expert and intuitively recognize which part should correct to mimic the expert's motion.



\begin{table}[]
\caption{Pearson's correlation between the joint difference of each body part and the distance in the latent space.}
\label{tab:corr}
\begin{tabular}{lclc}
Joint & Correlation & Joint & Correlation \\ \hline
Whole body & \multicolumn{1}{c|}{0.523} & Wrist & 0.468 \\
Elbow & \multicolumn{1}{c|}{0.507} & Spine & 0.468 \\
Shoulder & \multicolumn{1}{c|}{0.477} & Knee & 0.460 \\
Neck & \multicolumn{1}{c|}{0.473} & Foot & 0.455 \\
Head & \multicolumn{1}{c|}{0.471} & Hip & 0.441
\end{tabular}
\end{table}

\section{EVALUATION}

As our video database, We use the GolfDB~\cite{GolfDB}, which consists of 1400 high-quality video clips of experienced golfers' swings from online video platforms. 
To verify the effectiveness of the application, we investigated whether the distance in the latent space can represent the discrepancy between two input video frames.

We calculated the Mean Per Joint Position Error (MPJPE) of the two estimated 3D skeletons and compared it with the corresponding distance in the latent space.
In the experiment, we observed the correlation between the distance in the latent space and the MPJPE. 
This means that when the distance in the latent space is small, the difference of the two 3D poses is small (Figure~\ref{fig:teaser2}.c); when the distance is large, 3D poses have more differences (Figure~\ref{fig:teaser2}.b and d). 
Furthermore, when digging into human body parts, we found that particular body parts affect the distance in the latent space more than other body parts.
As shown in Table~\ref{tab:corr}, upper body joints are more correlative to the distance in the latent space than lower body joints.
Based on the results, we observed that the proposed tool can retrieve motion features that help the user target the key moves that they have to correct in priority.

As depicted in Figure~\ref{fig:teaser2}.a, we also observed that even in certain circumstances when the two skeletons appear similar, the distance in the latent space remains large. 
This may be explained by the fact that the network is inferring the difference of not only the human motion but also features outside the human body. 
In this case, the pose of the golf club, which is also considered critical during the swing, might be the main factor causing the large distance in the latent space.

\section{CONCLUSION}
In this work, we create the application for analyzing and visualizing the discrepancy of two input golf swing motions. The user can easily grasp the difference between the swings of various experts and their own during self-training.
Through our early experiment, it was confirmed that the proposed tool can detect discrepancies between two swing videos.
Furthermore, we give interpretation about the network that can help understand the correlation between the learned latent space and motion of the human body as well as the other props. 
In the future, we plan to implement the system that proposes optimal correction ways with analyzed factors that directly affect the user's golf swing performance.

\bibliographystyle{ACM-Reference-Format}
\bibliography{sample-base}


\begin{thebibliography}{4}


\ifx \showCODEN    \undefined \def \showCODEN     #1{\unskip}     \fi
\ifx \showDOI      \undefined \def \showDOI       #1{#1}\fi
\ifx \showISBNx    \undefined \def \showISBNx     #1{\unskip}     \fi
\ifx \showISBNxiii \undefined \def \showISBNxiii  #1{\unskip}     \fi
\ifx \showISSN     \undefined \def \showISSN      #1{\unskip}     \fi
\ifx \showLCCN     \undefined \def \showLCCN      #1{\unskip}     \fi
\ifx \shownote     \undefined \def \shownote      #1{#1}          \fi
\ifx \showarticletitle \undefined \def \showarticletitle #1{#1}   \fi
\ifx \showURL      \undefined \def \showURL       {\relax}        \fi
\providecommand\bibfield[2]{#2}
\providecommand\bibinfo[2]{#2}
\providecommand\natexlab[1]{#1}
\providecommand\showeprint[2][]{arXiv:#2}

\bibitem[\protect\citeauthoryear{{Dwibedi}, {Aytar}, {Tompson}, {Sermanet}, and
  {Zisserman}}{{Dwibedi} et~al\mbox{.}}{2019}]%
        {TCC}
\bibfield{author}{\bibinfo{person}{D. {Dwibedi}}, \bibinfo{person}{Y. {Aytar}},
  \bibinfo{person}{J. {Tompson}}, \bibinfo{person}{P. {Sermanet}}, {and}
  \bibinfo{person}{A. {Zisserman}}.} \bibinfo{year}{2019}\natexlab{}.
\newblock \showarticletitle{Temporal Cycle-Consistency Learning}. In
  \bibinfo{booktitle}{\emph{2019 IEEE/CVF Conference on Computer Vision and
  Pattern Recognition (CVPR)}}. \bibinfo{pages}{1801--1810}.
\newblock
\urldef\tempurl%
\url{https://doi.org/10.1109/CVPR.2019.00190}
\showDOI{\tempurl}


\bibitem[\protect\citeauthoryear{{Martinez}, {Hossain}, {Romero}, and
  {Little}}{{Martinez} et~al\mbox{.}}{2017}]%
        {3dposebaseline}
\bibfield{author}{\bibinfo{person}{J. {Martinez}}, \bibinfo{person}{R.
  {Hossain}}, \bibinfo{person}{J. {Romero}}, {and} \bibinfo{person}{J.~J.
  {Little}}.} \bibinfo{year}{2017}\natexlab{}.
\newblock \showarticletitle{A Simple Yet Effective Baseline for 3d Human Pose
  Estimation}. In \bibinfo{booktitle}{\emph{2017 IEEE International Conference
  on Computer Vision (ICCV)}}. \bibinfo{pages}{2659--2668}.
\newblock
\urldef\tempurl%
\url{https://doi.org/10.1109/ICCV.2017.288}
\showDOI{\tempurl}


\bibitem[\protect\citeauthoryear{{McNally}, {Vats}, {Pinto}, {Dulhanty},
  {McPhee}, and {Wong}}{{McNally} et~al\mbox{.}}{2019}]%
        {GolfDB}
\bibfield{author}{\bibinfo{person}{W. {McNally}}, \bibinfo{person}{K. {Vats}},
  \bibinfo{person}{T. {Pinto}}, \bibinfo{person}{C. {Dulhanty}},
  \bibinfo{person}{J. {McPhee}}, {and} \bibinfo{person}{A. {Wong}}.}
  \bibinfo{year}{2019}\natexlab{}.
\newblock \showarticletitle{GolfDB: A Video Database for Golf Swing
  Sequencing}. In \bibinfo{booktitle}{\emph{2019 IEEE/CVF Conference on
  Computer Vision and Pattern Recognition Workshops (CVPRW)}}.
  \bibinfo{pages}{2553--2562}.
\newblock
\urldef\tempurl%
\url{https://doi.org/10.1109/CVPRW.2019.00311}
\showDOI{\tempurl}


\bibitem[\protect\citeauthoryear{{Sun}, {Xiao}, {Liu}, and {Wang}}{{Sun}
  et~al\mbox{.}}{2019}]%
        {Hrnet}
\bibfield{author}{\bibinfo{person}{K. {Sun}}, \bibinfo{person}{B. {Xiao}},
  \bibinfo{person}{D. {Liu}}, {and} \bibinfo{person}{J. {Wang}}.}
  \bibinfo{year}{2019}\natexlab{}.
\newblock \showarticletitle{Deep High-Resolution Representation Learning for
  Human Pose Estimation}. In \bibinfo{booktitle}{\emph{2019 IEEE/CVF Conference
  on Computer Vision and Pattern Recognition (CVPR)}}.
  \bibinfo{pages}{5686--5696}.
\newblock
\urldef\tempurl%
\url{https://doi.org/10.1109/CVPR.2019.00584}
\showDOI{\tempurl}


\end{thebibliography}
\end{document}